\documentclass[preprints,article,accept,moreauthors,pdftex]{Definitions/mdpi} 
\firstpage{1} 
\pubvolume{xx}
\issuenum{1}
\articlenumber{1}
\pubyear{2020}
\copyrightyear{2020}
\history{This version: \today; Accepted: 2020-12-12, Fluids ISSN 2311-5521}

\usepackage{epsfig}
\usepackage[percent]{overpic}

\newcommand{\dd}{\partial}
\newcommand{\de}{{\rm \, d}}

\renewcommand{\vec}[1]{\mathbf{#1}}

\renewcommand{\P}{$P\ $}

\usepackage{psfrag}

\usepackage{xspace}
\providecommand{\MD}{\textbf{MD}\xspace}
\providecommand{\FD}{\textbf{FD}\xspace}

\allowdisplaybreaks
\usepackage{color}
\newcommand{\rs}[1]{#1}

\Title{Effects of Shell Thickness on Cross-Helicity Generation in
  Convection-Driven Spherical Dynamos}

\Author{Luis Silva\orcidC, Parag Gupta, David MacTaggart\orcidB{} and Radostin D. Simitev*\orcidA{}}

\AuthorNames{Luis Silva, Parag  Gupta, David MacTaggart and Radostin  D. Simitev}

\address{%
School of Mathematics and Statistics, University of Glasgow -- Glasgow G12 8QQ, UK}

\corres{Correspondence: Radostin.Simitev@glasgow.ac.uk}

\abstract{
The relative importance of the helicity and cross-helicity
electromotive dynamo effects for self-sustained magnetic field
generation by chaotic thermal convection in rotating spherical
shells is investigated as a function of shell thickness.
Two distinct branches of dynamo solutions are found to coexist in
direct numerical simulations for shell aspect ratios between 0.25 and
0.6 -- a mean-field dipolar regime and a fluctuating dipolar regime.
The properties characterising the coexisting dynamo attractors are
compared and contrasted, including differences in temporal behavior and  
spatial structures of both the magnetic field and rotating thermal
convection.
The helicity $\alpha$-effect and the cross-helicity $\gamma$-effect
are found to be comparable in intensity within the fluctuating dipolar
dynamo regime, where their ratio does not vary significantly with the
shell thickness. In contrast, within the mean-field dipolar dynamo regime
the helicity $\alpha$-effect dominates by approximately two orders of
magnitude and becomes stronger with decreasing shell thickness.
}

\keyword{rotating thermal convection, convection-driven dynamos,
numerical simulations, bistability, mean-field magnetohydrodynamics,
spherical shells} 

\begin{document}

\section{Introduction}

\looseness=-1
Thermal flows give rise to some of the most characteristic large-scale
features of cosmic objects -- their self-sustained magnetic fields
\cite{Parker1979,Brun2017}. 
For instance, the Sun and several of the planets in the Solar System
display substantial magnetic fields \cite{Zwaan1987,Busse2015}. The
solar magnetic field drives solar activity and strongly affects
planetary atmospheres \cite{Usoskin2017,Owens2013}. Earth’s field
shields life from solar  radiation \cite{Russell1991}. Farther out,
the gas giants, the ice giants, and the Jovian moons all have
significant magnetic fields \cite{Simitev2007a}.
These fields are sustained by dynamo processes in the interiors
or the atmospheres of their celestial hosts where vigorous
convective motions of electrically conductive fluids generate
large-scale electric currents \cite{Larmor1919,Moffatt1978,Charbonneau2014}.  
The convective flows are driven primarily by thermal buoyancy forces
due to thermonuclear fusion in stellar interiors and secular cooling
in planetary interiors, respectively.    
Thermal convection in celestial bodies is highly turbulent in nature
and, at the same time, strongly influenced both by rotation and by the
self-generated magnetic fields. Considerable attention has therefore
been devoted to 
this fascinating and important subject, and for topical reviews
we refer to the papers by \citet{Simitev2005b}, 
\citet{Jones2015}, \citet{Wicht2019} and references
within.

\looseness=-1
Conceptually, dynamo generation of large-scale magnetic fields is  
understood on the basis of the mean-field dynamo theory
\cite{KrauseRaedler1980,Brandenburg2018,Moffatt2019}, a
well-established theory of magnetohydrodynamic turbulence. 
A cornerstone of the theory is the turbulence modelling of the mean
electromotive  force -- the sole source term arising in the Reynolds-averaged magnetic
induction equation governing the evolution of the large-scale field, see Section
\ref{sec:theCHEffect} further below.  The electromotive force is usually
approximated by an expansion in terms of the mean field and its
spatial derivatives where the expansion coefficients are known informally
as ``mean-field effects''. 
The turbulent helicity effect\footnote{In this work, when we refer to
``helicity'' without further qualification, we intend the helicity
associated with the $\alpha$-effect. This shorthand should not be
confused with other helicities, such as ``magnetic helicity''.}, 
also called  \emph{$\alpha$--effect}, has been
studied extensively in the research literature on mean-field dynamo theory,
e.g.~see \cite{Brandenburg2018,Brandenburg2005} and references
therein. 
In contrast, the cross-helicity effect, also known as
\emph{$\gamma$--effect} \cite{Yoshizawa1993}, has been a subject to a
rather small number of studies, e.g.~\cite{Yokoi2013,Pipin2018} and works cited
therein. This is due to the currently prevailing treatment of 
turbulence where large-scale velocity is neglected because of the Galilean
invariance of the momentum equation. However, such treatment leads to the
neglect of the large-scale shear effects which are, in fact,
significant. For example, large-scale rotation is  ubiquitous in 
astro/geophysical objects, e.g.~the Solar internal differential 
rotation is substantial and well measured 
\cite{Thompson1996,Schou1998} while numerical simulations
suggest it is an essential ingredient of the dynamo process and likely
to be responsible for the regular oscillations of convection-driven
spherical dynamos \citep{Simitev2006b,Simitev2012a}.  
Similarly, a number of  studies of plane-parallel flows confirm
that cross-helicity effects are not small compared to helicity effects
\citep{Hamba1992,Yokoi2011}.  Apart from its role in dynamo
generation, cross-helicity is an important Solar observable. For
instance, measurements of the cross-helicity component $\langle u_z
b_z\rangle$ at the Solar surface are available from the Swedish 1-m
Solar Telescope  and can be used to calculate the magnetic eddy
diffusivity of the quiet Sun by quasilinear mean-field theory \citep{Ruediger2012}.

\begin{figure}[t]
\begin{center}
\begin{overpic}[height=2cm,width=12cm]{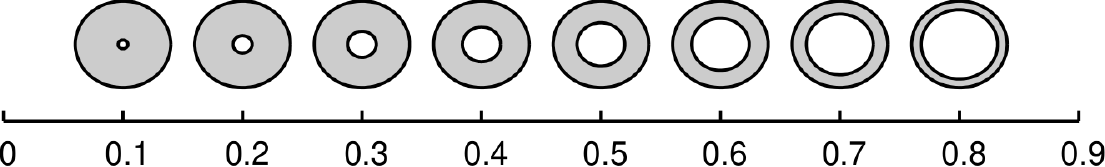}
\put (48,-2) {{$\eta$}}
\end{overpic}
\end{center}
\caption[]{Illustration of shell thickness aspect ratio variation.}
\label{fig:cartoon}
\end{figure}

Cross-helicity has not been explored in models of self-consistent
dynamos driven by thermal convection in rotating spherical shells and
this paper aims to contribute in this direction. The main goal of this work
is to investigate the relative importance of the helicity 
and cross-helicity effects as a function of the thickness of the convective shell.
Intuitive arguments suggest that the $\alpha$--effect is important in the case
of the geodynamo and the cross-helicity effect is important in the
case of the global solar dynamo. Indeed, the geodynamo operates in the
relatively thick fluid outer code of the Earth where large-scale
columnar structures are believed to develop. The coherent columnar
structures are characterised by relatively large-scale vorticity and
generate a strong helicity $\alpha$--effect. In contrast, the global
solar dynamo operates in the thinner solar convection zone where
columnar structures are thought difficult to maintain and 
so vorticity may have a less regular structure, thus increasing the relative importance of
the cross-helicity effect. To assess this hypothesis, we present a set of dynamo
simulations that differ mainly in their shell thickness aspect ratio
$\eta = r_i/r_o$, see Figure \ref{fig:cartoon}, while other governing
parameters are kept fixed. Along with estimates of the relative strength of the helicity and cross-helicity 
effects, we report on the mechanisms of electromotive force generation and its
spatial distribution. Variation of shell thickness is also relevant
to the case of the geodynamo as the inner core did not exist at the
time of formation of the Earth, but nucleated sometime later in the
geological history of the planet and continues to grow in size.

The geodynamo and the solar global dynamo are also different in that
the former has a dominant and rarely reversing dipole, while the
latter exhibits a regular periodic cycle. To capture this essential
difference while comparing quid pro quo, we have performed this study
at parameter values where two distinct dynamo branches are known to coexist
\cite{Simitev2009,Simitev2011b,Simitev2012c}.  These branches have
rather different magnetic field properties, in particular one branch is
non-reversing while the other branch is cyclic, and also display significant
differences in zonal flow intensity and profile. It is reasonable to expect 
that the two branches will offer different mechanisms of helicity and
cross-helicity generation and thus in this paper we proceed to
study both branches. Bistability, in itself, may play a role in aperiodic
magnetic field polarity reversals, a notable feature of the geodynamo
\cite{Simitev2008a}, as well as in the regular cycle of the solar
dynamo \cite{Matilsky2020}. 
We have previously investigated the hysteretic transitions between the
coexisting dynamo branches with variation of the Rayleigh, Prandtl and
Coriolis numbers (defined further below). In addition, in this paper
we demonstrate for the first time that the distinct dynamo branches
coexist also when the shell thickness $\eta$ is varied. The discussion
of this dichotomous behaviour runs as a secondary theme of the article.  

The paper is structured as follows.
Details of the mathematical model and the numerical methods for
solution are given in section \ref{ap:math}.
In section \ref{sec:res}, we describe the set of dynamo simulations
performed in the context of this work. We pay particular attention to
the description of the two coexisting dynamo branches which are
studied for the first time here as a function of the thickness of the
convective shell. In the process, we describe the typical morphology
and time dependent behaviour of thermal convection flows. In section
\ref{sec:theCHEffect}, we briefly summarise the mean field arguments
related to the helicity and cross-helicity mechanisms for the generation
of large-scale magnetic field. In section \ref{sec:effectsImportance},
the cross-helicity properties of our dynamo solutions and the relative
contributions of the $\alpha$-- and $\gamma$--effects are
assessed. Section \ref{sec:disc} is devoted to concluding remarks.   

\begin{figure}
\begin{center}
\hspace*{10mm}
\includegraphics[height=6cm]{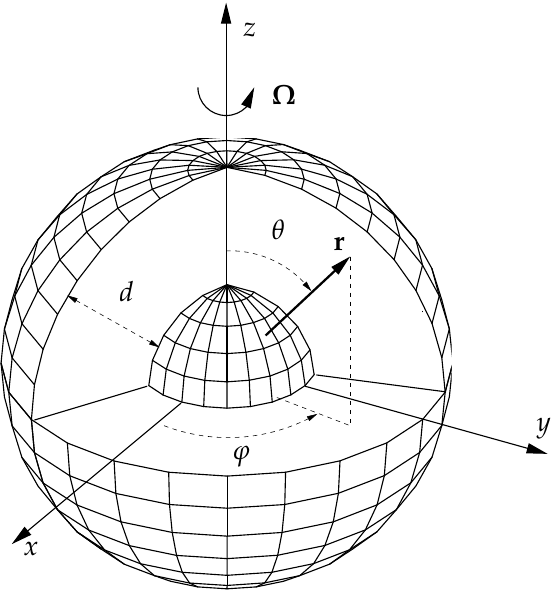}
\end{center}
\caption[]{\rs{Schematic illustration of the three-dimensional region
    considered in this study, the associated spherical coordinate
    system and the position of the axis of rotation. The region is
    assumed full of electrically conducting fluid.}}
\label{fig:geometricalSetUp}
\end{figure}

\section{Materials and Methods}
\label{ap:math}
\unskip

This section describes a standard mathematical formulation of the
problem of thermal convection and magnetic field generation in
rotating spherical fluid shells. A set of transformations used to
recast the problem in a scalar stream-function form and a
pseudo-spectral algorithm used for numerical solution of the equations
is presented. 
The exposition in this section is standard and follows our
previous articles, e.g.~\cite{Simitev2005a,Mather2020}.
This section also serves as an introduction and a review of the typical
approach to the formulation and solution of this important  problem.

\subsection{Mathematical formulation}
\label{formulation}
We consider a spherical shell full of electrically conducting
fluid as shown in Figure \ref{fig:geometricalSetUp}.
The shell rotates with a constant angular velocity $\vec{\Omega}$
about the vertical coordinate axis. 
We assume that a static state exists with the temperature distribution
\begin{subequations}
\begin{gather}
T_S = T_0 - \beta d^2 r^2 /2, \\
\beta=q/(3\,\kappa\,c_p),\\
T_0=T_1-(T_2-T_1)/(1-\eta).
\end{gather}
\end{subequations}
\begin{table}[t]
\caption{Notation used in section \ref{formulation}, where not defined
  in the main text.}
\label{table}
\begin{tabular}{cp{0.35\textwidth}cp{0.35\textwidth}}
\toprule
\textbf{Notation}     & \textbf{Quantity}  &  \textbf{Notation}	& \textbf{Quantity} 	\\
\midrule
$(r, \theta, \varphi)$ & Spherical polar coordinates &
$T_S$ & Background temperature distribution \\
$t$ & Time &
$T_1,T_2$ & Temperature inner, outer  boundary \\
$\vec r$ & Position vector wrt centre of sphere	&
$q$ 	& Density of uniformly distributed  heat sources\\
$d$ & Thickness of the spherical shell &
$\kappa$ & Thermal diffusivity \\
$r_i$, $r_o$ &  Inner and outer radii of the shell &
$\nu$ 	& Kinematic viscosity\\
$\vec u$  & Velocity field perturbation	&
$\mu$	& Magnetic permeability \\
$\vec B$  & Magnetic flux density perturbation &
$c_p$ & Specific heat at constant pressure\\
$\Theta$  & Temperature perturbation from the background state &
$\gamma$ & Gravitational acceleration magnitude	 \\
$\pi$ & Effective pressure &
$\partial$	& Partial derivative notation\\
\bottomrule
\end{tabular}
\end{table}
\rs{The evolution of the system is governed by the equations of
momentum, heat and magnetic induction, along with solenoidality
conditions for the velocity and magnetic fields,}
\begin{subequations}
\begin{gather}
\label{1b}
\nabla \cdot \vec u = 0, \\
\label{1a}
\big(\partial_t + \vec u \cdot \nabla\big) \vec u = - \nabla \pi -  \tau
\vec k \times\vec u +\Theta \vec r + \nabla^2 \vec u + \vec B \cdot \nabla \vec B, \\
\label{1c}
P\big(\partial_t + \vec u \cdot \nabla \big)\Theta = R \vec r \cdot \vec u + \nabla^2 \Theta, \\
\nabla \cdot \vec B = 0, \\
\label{1d}
P_m\big(\partial_t + \vec u \cdot \nabla \big) \vec B =
P_m  \vec B \cdot \nabla \vec u + \nabla^2 \vec B,
\end{gather}
\end{subequations}
\rs{written for the perturbations from the static reference state and
  with notations defined in Table \ref{table}. In this formulation,
  the Boussinesq approximation is used with the density
$\varrho$ having a constant value $\varrho_0$ except in the gravity term where}
\begin{gather}
\varrho = \varrho_0(1 - \alpha\Theta),
\end{gather}
and $\alpha$ is the specific thermal expansion coefficient $\alpha
\equiv - ( \de \varrho/\de T)/\varrho = \text{const}$. 
With the units of Table \ref{tableunits}, five dimensionless parameters appear in the governing equations,
namely the shell radius ratio $\eta$, the Rayleigh number $R$, 
the Coriolis number $\tau$, the Prandtl number $P$ and the magnetic
Prandtl number $P_m$  defined by 
\begin{equation}
\eta=\frac{r_i}{r_o}, \enspace R = \frac{\alpha \gamma \beta d^6}{\nu \kappa} , 
\enspace \tau = \frac{2
\Omega d^2}{\nu} , \enspace P = \frac{\nu}{\kappa} , \enspace P_m = \frac{\nu}{\lambda},
\end{equation}
where $\lambda$ is the magnetic diffusivity. Since the velocity 
$\vec u$ and the magnetic flux density $\vec B$ are
solenoidal vector fields,   the general representation in terms of
poloidal and toroidal components is used
\begin{subequations}
\begin{align}
&
\vec u = \nabla \times ( \nabla v \times \vec r) + \nabla w \times 
\vec r, \\
&
\vec B = \nabla \times  ( \nabla h \times \vec r) + \nabla g \times 
\vec r.
\end{align}
\end{subequations}
Taking $\vec r\cdot \nabla\times$ and $\vec r\cdot \nabla\times \nabla\times$ 
of the momentum equation \eqref{1a}, two equations for $w$ and $v$ are obtained
\begin{subequations}
\label{ScalEQs}
\begin{gather}
\label{momentumw}
[( \nabla^2 - \partial_t) {\cal L}_2 + \tau \partial_{\varphi} ] w - \tau {\cal Q}v 
= \vec
r \cdot \nabla \times ( \vec u \cdot \nabla \vec u - \vec B \cdot
\nabla \vec B),\\
\label{momentumv}
[( \nabla^2 - \partial_t) {\cal L}_2 + \tau \partial_{\varphi} ] \nabla^2 v +
\tau {\cal Q} w - {\cal L}_2 \Theta  
= - \vec r \cdot \nabla \times [ \nabla \times ( \vec u \cdot
\nabla \vec u - \vec B \cdot \nabla \vec B)],
\end{gather}
where $\partial_{\varphi}$ denotes the partial derivative with respect to
the angle $\varphi$ of a spherical system of coordinates $(r, \theta, \varphi)$
and where the operators ${\cal L}_2$ and $\cal Q$ are defined as
\begin{gather}
{\cal L}_2 \equiv - r^2 \nabla^2 + \partial_r ( r^2 \partial_r), \nonumber\\
{\cal Q} \equiv r \cos \theta \nabla^2 - ({\cal L}_2 + r \partial_r ) ( \cos \theta
\partial_r - r^{-1} \sin \theta \partial_{\theta}). \nonumber
\end{gather}
The heat equation for the dimensionless deviation $\Theta$ from the
static temperature distribution can be written in the form
\begin{equation}
\label{heat}
\nabla^2 \Theta + R{\cal L}_2 v = P ( \partial_t + \vec u \cdot \nabla ) \Theta,
\end{equation}
and the equations for $h$ and $g$ are obtained by taking $\vec r\cdot$
and $\vec r\cdot \nabla\times$  of the dynamo equation
\eqref{1d} 
\begin{gather}
\label{inductionh}
\nabla^2 {\cal L}_2 h = P_m [ \partial_t {\cal L}_2 h - \vec r \cdot
\nabla \times ( \vec u \times \vec B )], \\
\label{inductiong}
\nabla^2 {\cal L}_2 g = P_m [ \partial_t {\cal L}_2 g - \vec r \cdot
\nabla \times ( \nabla \times ( \vec u \times \vec B ))].
\end{gather}
\end{subequations}

\begin{table}[t]
\caption{Units of non-dimensionalisation.}
\centering
\label{tableunits}
\begin{tabular}{p{0.2\textwidth}l}
\toprule
\textbf{Quantity}     & \textbf{Unit} \\
\midrule
Length & $d$ \\
Time &  $d^2 / \nu$\\
Temperature & $\nu^2 / \gamma \alpha d^4$ \\
Magnetic flux density & $\nu ( \mu \varrho )^{1/2} /d$ \\
\bottomrule
\end{tabular}
\end{table}

For the flow we assume stress-free boundaries with fixed temperatures 
\begin{subequations}
\label{BCs}
\begin{gather}
\label{vbc}
v = \partial^2_{rr}v = \partial_r (w/r) = \Theta = 0 
\quad \mbox{ at } r=r_i \mbox{ and } r=r_o.
\end{gather}
For the magnetic field we assume electrically insulating
boundaries such that the poloidal function $h$ must be 
matched to the function $h^{(e)}$ which describes the  
potential fields outside the fluid shell  
\begin{gather}
\label{mbc}
g = h-h^{(e)} = \partial_r ( h-h^{(e)})=0 
\quad \mbox{ at } r=r_i \mbox{ and } r=r_o.
\end{gather}
\end{subequations}

This is a standard formulation of the spherical convection-driven dynamo problem
\cite{Busse2000,Dormy2007,Roberts2013,Jones2015} for which an extensive
collection of results already exists \cite{Grote2000,Simitev2003,Simitev2005a,Simitev2006b}.
The results reported below are not strongly model dependent as
confirmed by simulations of convection driven by differential heating
\cite{Simitev2003b}, for cases with no-slip conditions at the
inner boundary and an electrical conductivity of the exterior equal to
that of the fluid \cite{Simitev2012a,Simitev2012b}, and for
thermo-compositional driving \cite{Mather2020}. Thus, aiming to retain
a general physical perspective, we intentionally use here a generic
model formulation with a minimal number of physical parameters
including only those of first-order importance for stellar and planetary
applications.  

\subsection{Numerical methods}

For the direct numerical integration of the convection-driven dynamo
problem specified by the scalar equations \eqref{ScalEQs} and the
boundary conditions \eqref{BCs} we use a pseudo-spectral method
described by \cite{Tilgner1999}. The code has been benchmarked for
accuracy, most recently in \citep{Marti2014,Matsui2016}, and has been
made open source \citep{Silva2018b}.  
\rs{All dependent variables in the code are spatially discretised by
  means of spherical harmonics ${\mathrm Y}_l^m$ and Chebychev polynomials $T_n$, e.g.} 
\begin{equation}
\label{6}
v(r,\theta, \varphi) = \sum \limits_{l,m,n}^{N_l,N_m,N_n} V_{l,n}^m
(t) T_n\big(2(r-r_i)-1\big) {\mathrm Y}_l^m (\theta, \varphi),
\end{equation}
\rs{and similarly for the other unknown scalars, $w$, $h$, $g$ and
  $\Theta$. The nonlinear terms in
  the equations are computed in physical space and then projected onto
  spectral space at every time step. Time integration makes use of an
  IMEX combination of the Crank-Nicolson scheme for the diffusion terms and
  the Adams-Bashforth scheme for the nonlinear terms, both schemes of
  second order accuracy.}

\rs{When the spectral powers of the kinetic and magnetic energies drop
  by more than three orders of magnitude from the spectral maximum to
  the cut-off wavelength, we consider the simulations to be reasonably
  resolved \cite{Christensen1999}. In all the cases reported here, a
  minimum of 41 collocation points in the radial direction has been
  considered, together with spherical harmonics up to order 96. These
  numbers provide sufficient resolution, as demonstrated in
  Figure~\ref{fig:exampleSpectra} for two typical dynamo solutions.} 

\subsection{Diagnostics}
It is convenient to characterise the non-magnetic convection and
the convection-driven dynamo solutions using their energy densities. To
understand the interactions between various components of the flow, we
decompose the kinetic energy density into 
mean poloidal, mean toroidal, fluctuating  poloidal and fluctuating toroidal parts as follows
\begin{subequations}
\begin{gather}
\overline{E}_p = \frac{1}{2} \langle \mid\nabla \times ( \nabla \overline{v}
\times \vec r )\mid^2 \rangle ,  \quad
 \overline{E}_t = \frac{1}{2} \langle \mid\nabla
\overline w \times \vec r \mid^2 \rangle, \\
\widetilde{E}_p = \frac{1}{2} \langle \mid\nabla \times ( \nabla \widetilde v
\times \vec r )\mid^2 \rangle , \quad
 \widetilde{E}_t = \frac{1}{2} \langle \mid\nabla
\widetilde w \times
\vec r \mid^2 \rangle,
\end{gather}
\end{subequations}
where $\langle\cdot\rangle$ indicates the average over the fluid shell
and time as described in section \ref{sec:theCHEffect} and  
$\overline v$ refers to the axisymmetric component of the poloidal
scalar field $v$, while $\widetilde v$ 
is defined as $\widetilde v = v - \overline v$. The corresponding magnetic
energy densities $\overline{M}_p$, $\overline{M}_t$, $\widetilde{M}_p$ and 
$\widetilde{M}_t$ are defined analogously with the scalar fields $h$
and $g$ for the magnetic field replacing $v$ and $w$.

To assess the predominant configuration of the magnetic field, we
define the dipolarity ratio
\begin{gather}
{\cal D} = \overline{M}_p/\widetilde{M}_p.
\end{gather}
When $\overline{M}_p > \widetilde{M}_p$ then ${\cal D} > 1$ and the corresponding 
solutions will be referred to as ``Mean Dipolar'', for reasons to be
explained below, and denoted by \MD{} following
\cite{Simitev2009}.
When $\overline{M}_p < \widetilde{M}_p$ then ${\cal D} < 1$ and the corresponding 
solutions will be referred to as ``Fluctuating Dipolar'' and denoted
by \FD{}.

To quantify heat transport by convection the Nusselt numbers at the inner and outer spherical boundaries $Nu_i$
and $Nu_o$ are used. These are  defined by 
\begin{equation}
\label{nu.def}
Nu_i=1- \frac{P}{r_iR} \left.\frac{\de \overline{\overline{\Theta}}}{\de r}\right|_{r=r_i},
 \qquad
Nu_o=1- \frac{P}{r_oR} \left.\frac{\de \overline{\overline{\Theta}}}{\de r}\right|_{r=r_o},
\end{equation}
where the double bar indicates the average over the spherical
surface.

Other quantities are defined in the text as required.

\begin{figure}[t]
\begin{center}
\includegraphics[width=\textwidth,clip=true]{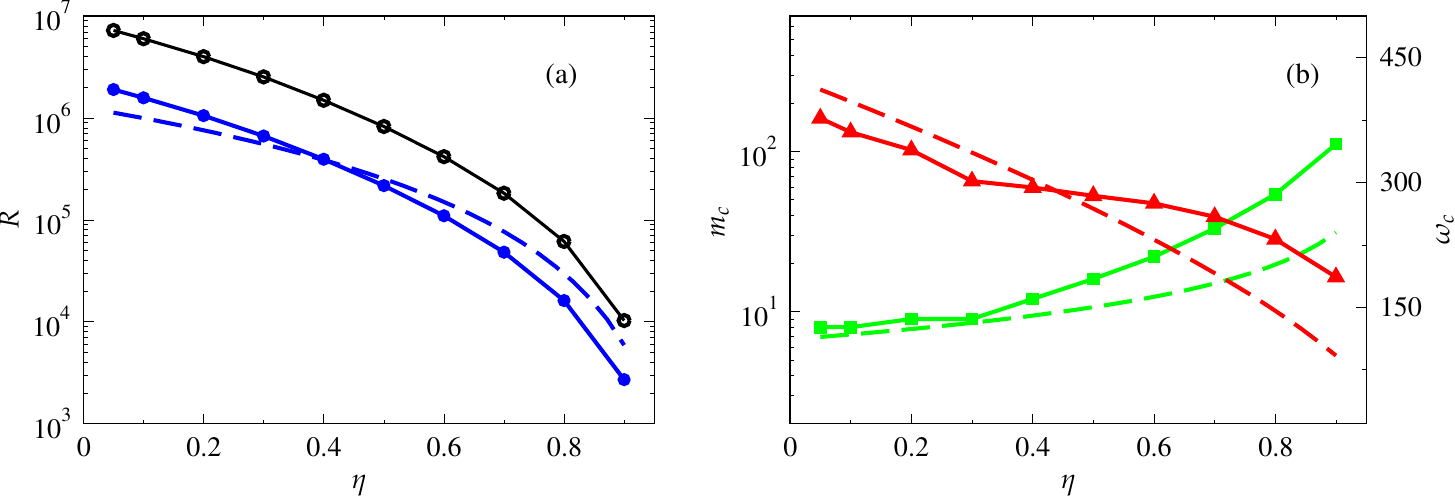}
\end{center}
\caption{
Critical parameter values for the onset of convection and values of
the Rayleigh number used in this work as a function of the  
shell thickness aspect ratio $\eta$ in the case $P=0.75$, and $\tau= 2 \times 10^4$. 
(a) The critical Rayleigh number $R_c$ for the linear onset of convection is 
plotted in solid blue curve marked by full circles. The values used in the 
simulations are given by $R=3.8 R_c$; they are plotted in solid black
curve marked by empty circles. (b) The critical wave number $m_c$ (left $y$-axis) and the critical 
frequency $\omega_c$ (right $y$-axis) for the onset of convection are denoted 
by green squares and red triangles, respectively. Local asymptotic
approximations \eqref{highPasym} are shown by correspondingly colored dashed
curves in all panels.
(color online)}
\label{fig:Rayleigh}
\end{figure}

\section{Results}
\label{sec:res}


\subsection{Parameter values used}
In order to investigate the effects of the shell thickness on the properties of
non-magnetic convection and on dynamo solutions we perform a suite
of numerical simulations varying the shell aspect ratio between $\eta= 0.1$
and $\eta=0.7$. To compare the simulations on equal footing, as  well
as to keep the number of runs required to a manageable  
level, all parameters except those depending on the aspect ratio are
kept at fixed values. The value of the Prandtl number is set to
$P=0.75$ allowing us to use a relatively low value of the
magnetic Prandtl  number $P_m=1.5$ as appropriate for natural
dynamos. The Coriolis number is fixed to $\tau=2 \times 10^4$ 
representing a compromise between the fast  rotation rate appropriate
for the geodynamo and the relatively slow rotation  rate appropriate
for the solar dynamo. To ensure that dynamos are driven  equally
strongly, we fix the value of the Rayleigh number at 3.8 times the
critical value $R_c$ for the onset of convection for each shell
thickness aspect ratio as shown in Figure \ref{fig:Rayleigh} below.
The required values of the critical Rayleigh number are determined as
explained in the next section where we also discuss general features
of the onset of thermal convection.
\begin{figure}[t]
\begin{center}
  \includegraphics[width=\textwidth,clip=true]{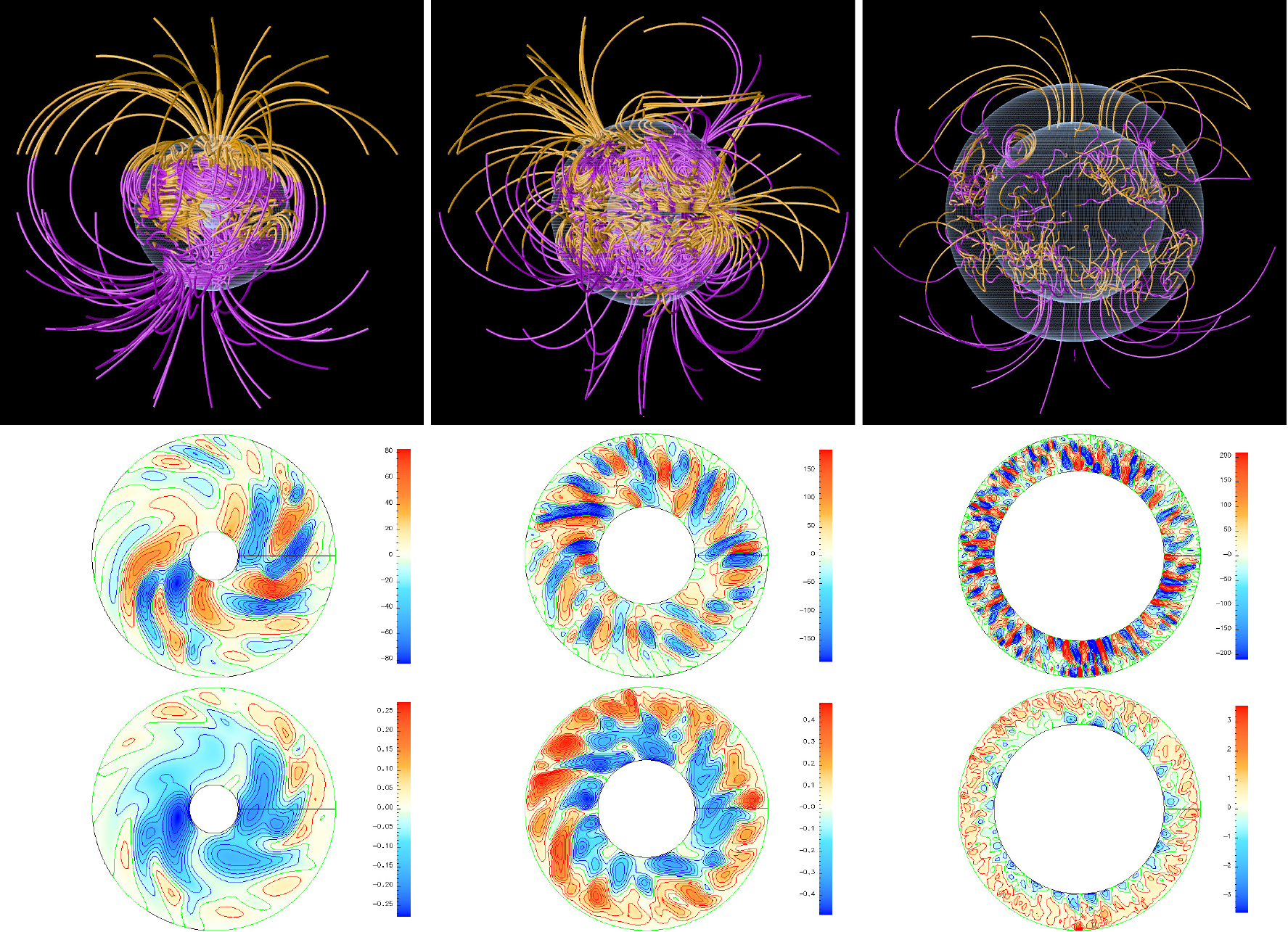}
\end{center}
\caption{Snapshots of spatial structures of dynamo solutions
  with increasing shell thickness aspect ratio $\eta$ and with
  $R=3.8 \times R_c$, $\tau=2\times10^4$, $P=0.75$ and $P_m=1.5$.
  Three cases are shown as follows: $\eta=0.2$, $R=4000000$ (left column);
  $\eta=0.4$, $R=1500000$ (middle column); and
  $\eta=0.7$, $R=180000$ (right column).
  Magnetic poloidal fieldlines are plotted in the top row,
  contours of the radial velocity $u_r$ in the equatorial plane are
  plotted in the middle row, and contours of the temperature
  perturbation $\Theta$ in the equatorial plane are plotted in the
  bottom row. 
  (color online)
}
\label{fig:3D}
\end{figure}

\subsection{Linear onset of thermal convection}

The onset of thermal convection in rapidly rotating spherical shells has been
extensively studied, e.g.~most recently as a special case of the onset
of thermo-compositional convection \cite{Silva2019}. In general, two major regimes are found
at onset -- columnar convection and equatorially-attached convection. 
The equatorially-attached regime occurs at small values of the Prandtl
number $P$ and consists of flows that take the form of non-spiralling rolls trapped
near the equator with a relatively large azimuthal length scale. This
regime can be understood as a form of inertial oscillations,
e.g~\cite{Simitev2004}. The columnar regime is realised at moderate and large values of $P$
and features elongated rolls parallel to axis of rotation that are
spiralling strongly and have a relatively short azimuthal length scale. At the selected values of the
Prandtl and the Coriolis numbers, the simulations reported in
this study belong to the columnar regime of rapidly rotating convection.

To determine accurate values for the critical parameters at onset we
use our open source numerical code \cite{Silva2018a}. The code
implements a Galerkin spectral projection method due to \citet{Zhang1987} to
solve the linearised versions of equations
(\ref{momentumw}--\ref{heat}). The method leads to a generalised
eigenvalue problem for the critical Rayleigh number $R_c$ and
frequency $\omega_c$ of
the most unstable mode of thermal convection at specified other
parameter values and at specified azimuthal wave number $m$ of the
convective perturbation. Numerical extremisation and continuation 
problems then are tackled in order to follow the marginal  stability
curve in the parameter space as detailed in \cite{Silva2019}.
The critical values thus obtained are shown in
Figure~\ref{fig:Rayleigh}. The critical Rayleigh number $R_c$ and drift
frequency $\omega_c$ decrease with decreasing shell thickness while
the critical azimuthal wave number $m_c$ increases.

\looseness=-1
It is interesting to compare {and validate} these results against
theoretical results for the onset convection in rapidly rotating
systems. The asymptotic analysis of this problem has a long and
distinguished history \rs{of local and global linear stability analysis
\cite{Roberts1968,Busse1970,Soward1977,JonesSoMu2000,DORMY2004}, see
also \cite{Silva2019} for a brief overview. 
Converting results of \citet{Yano1992} to our dimensionless
parameters, length and time scales, we obtain}
\begin{subequations}
\label{highPasym}  
\begin{gather}
R_c=7.252\left(\frac{P \tau}{1+P}\right)^{4/3} (1-\eta)^{7/3},\\
m_c=0.328\left(\frac{P \tau}{1+P}\right)^{1/3}(1-\eta)^{-2/3},
 \\
\omega_c=0.762\left(\frac{\tau^2}{P(1+P)^2}\right)^{1/3}(1-\eta)^{2/3},
\end{gather}
\end{subequations}
\rs{for the critical parameters of viscous columnar convection in an
internally heated spherical shell.}
{While expressions \eqref{highPasym} are not strictly valid asymptotic 
results for the spherical shell configuration studied here, they
provide a reasonable agreement with the numerical results
plotted in Figure~\ref{fig:Rayleigh}.
{While expressions \eqref{highPasym} are not strictly valid asymptotic 
results for the spherical shell configuration studied here, they
provide a reasonable agreement with the numerical results
plotted in Figure~\ref{fig:Rayleigh}.

\begin{figure}[t]
\begin{center}
\vspace*{-3.5mm}
\includegraphics[width=\textwidth,clip=true]{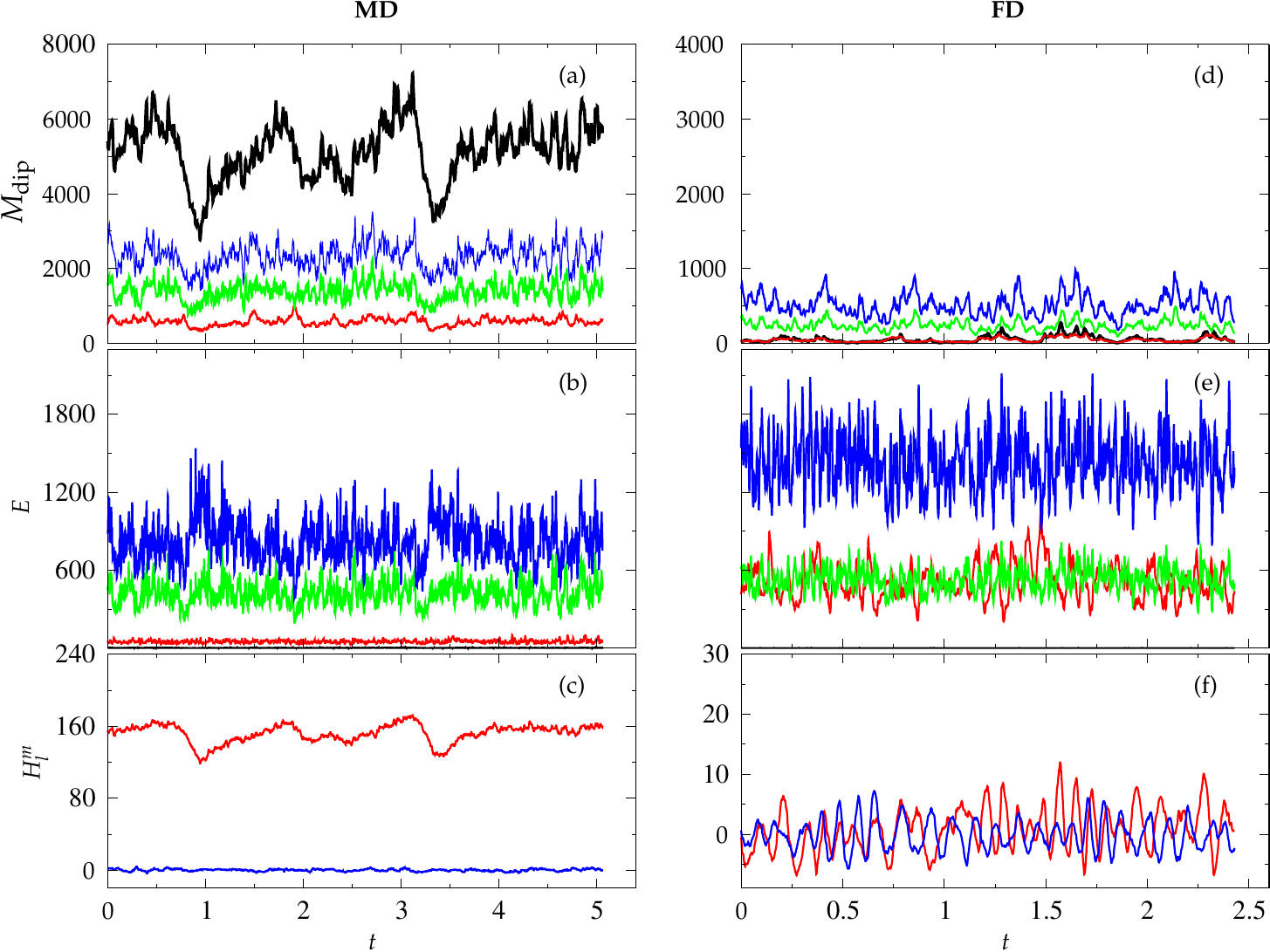}
\end{center}
\caption[]{
Chaotic dynamo attractors at identical parameter  
values -- a MD dynamo (left column (a,b,c)) and a FD dynamo (right column (d,e,f)) 
both at $\eta=0.5$, $R=8.2\times10^5$, $\tau=2\times10^4$, $P=0.75$ and  
$P_m=1.5$.  Panels (a,d) show time series of magnetic dipolar energy densities 
and panels (b,e) show kinetic energy densities. The component $\overline{X}_p$ 
is shown by solid black line, while $\overline{X}_t$, $\widetilde{X}_p$, and 
$\widetilde{X}_t$ are shown by red, green and blue lines, respectively. $X$ 
stands for either $M$ or $E$. Panels (c,f) show the axial dipolar $H_1^0$ and 
the axial quadrupolar $H_2^0$ coefficients at midshell $r=(r_i+r_o)/2$
by red and blue lines,  respectively. Note the very different ordinate
scales between panels (a) and  (d) and (c) and (f). The ordinate
scales of panels (b) and (e) are identical.
(color online)}
\label{fig:examplePlots}
\end{figure}

\subsection{Finite-amplitude convection and dynamo features}

\looseness=-1
As the value of the Rayleigh number is increased away from the onset,
rotating columnar convection undergoes a sequence of transitions from
steady flow patterns drifting with constant angular velocity to
increasingly chaotic states as described in detail in \cite{Simitev2003}.
When the amplitude of convection becomes sufficiently large so
that the magnetic Reynolds number defined as $Rm = Pm\sqrt{2E}$
reaches values of the order $10^2$, onset of dynamo action is typically
observed \cite{Simitev2005a}.
Three examples of dynamo solutions are shown in Figure \ref{fig:3D} to
(i) illustrate typical spatial features of chaotic thermal convection
in rotating shells and the associated magnetic field morphology and
(ii) to reveal how these features vary with decreasing shell thickness.
Outside of the tangent cylinder the flow consists of pairs of adjacent
spiralling convection columns as seen in the second row of Figure \ref{fig:3D}.
Within the columns the fluid particles travel in clockwise and
anticlockwise directions parallel to the equatorial plane and up
towards the poles or down towards the equatorial plane as columns
extend through the height of the convective shell. In agreement
with the linear analysis, as the shell thickness is decreased the
azimuthal wave number rapidly increases with the thin shell
solution $\eta=0.7$ showing a cartridge of fine scale columns closely
adjacent to each other and exhibiting much weaker spiralling and slower
drift than in the thick shell cases.
These convective patterns strongly influence the structure and the
morphology of magnetic fields as illustrated by the first row of
Figure \ref{fig:3D} where magnetic fieldlines of the three dynamo
solutions are shown. The fieldlines are intricately knotted and exhibit a 
rather complicated structure within the convective domain in all three
cases. The imprint of the convective columns is visible in the thick
shell cases $\eta=0.2$ and $\eta=0.4$ where the magnetic fieldlines
are coiled around the convective columnar structures indicating the
presence of toroidal field and poloidal field feedback and
amplification processes. Outside of the convective domain, the
magnetic field of the thickest shell case $\eta=0.2$ is well organized
and emerges from the polar regions of the domain in the form of big
bundles of opposite polarities with fieldlines proceeding to close 
and forming extensive overarching loops that are characteristic of a
strong dipolar field symmetry. A similar picture is seen in the mid-thickness
case $\eta=0.4$ although in this case there appear to be several
magnetic ``poles'' where strong bundles of vertical fieldlines emerge
at the surface of the spherical domain. In the thin shell case
$\eta=0.7$ the magnetic field is much less organized with numerous
{fieldline} coils inside the convective domain and barely visible but
still dominant dipolar structure outside.
\begin{figure}[t]
\begin{center}
\includegraphics[width=\textwidth,clip=true]{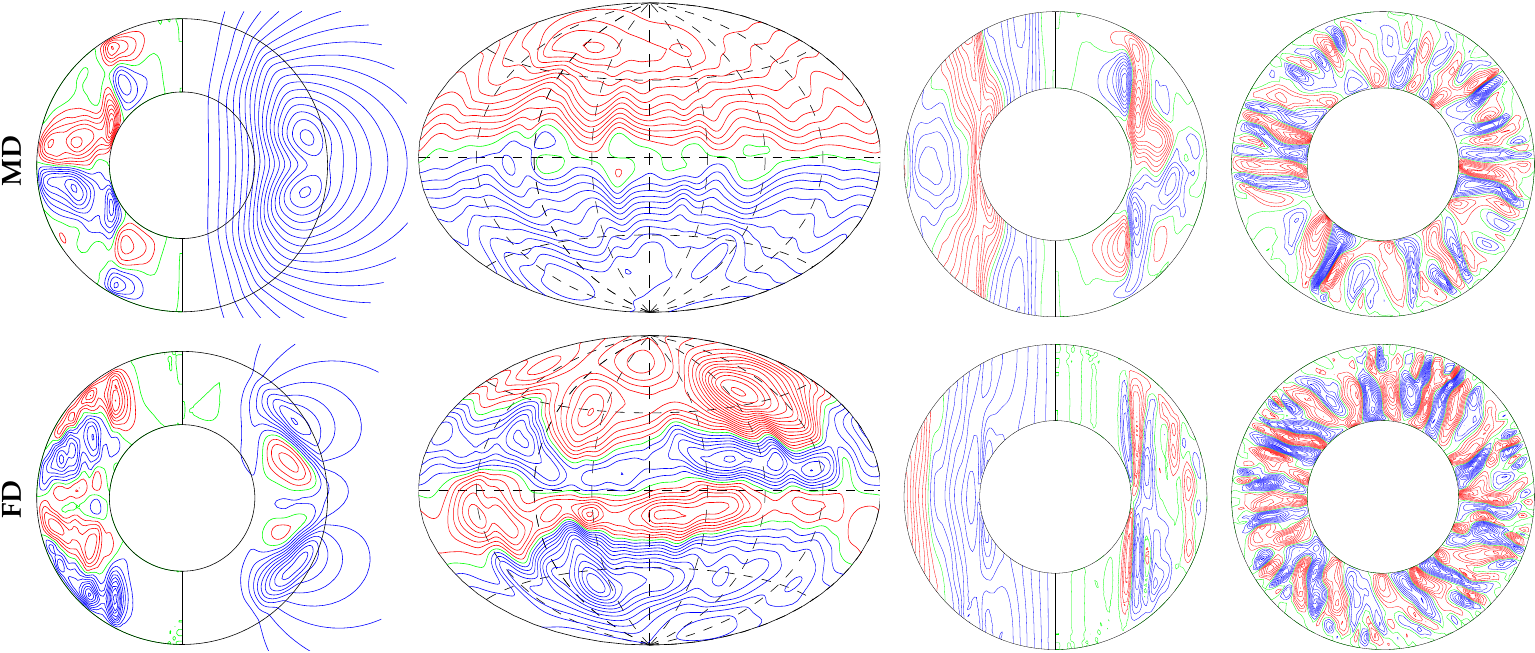}
\end{center}
\caption{
A \MD{} (top row) and a \FD{} (bottom row) dynamo 
solutions at $\eta=0.5$, $R=8.2\times 10^5$, $\tau=2\times10^4$, $P=0.75$ and  
$P_m=1.5$ corresponding to the cases shown in Figure \ref{fig:examplePlots}. 
The first column shows meridional lines of constant $\overline{B_{\varphi}}$ in the 
left half and of $r \sin \theta \dd_\theta \overline{h}=const.$ in the right 
half. The second column shows lines of constant $B_r$ at $r=1.675
r_o$. The third column 
shows meridional lines of constant $\overline{u}_\varphi$ in the left half and 
of $r \sin \theta \dd_\theta \overline{v}$ in the right half. The fourth column 
shows contours of the radial flow $u_r$ on the equatorial plane. Positive values are 
shown in red; negative values are shown in blue, and the zeroth
contour line is shown in green. 
(color online)}
\label{fig:exampleMaps}
\end{figure}
\begin{figure}[t]
\begin{center}
\includegraphics[width=\textwidth,clip=true]{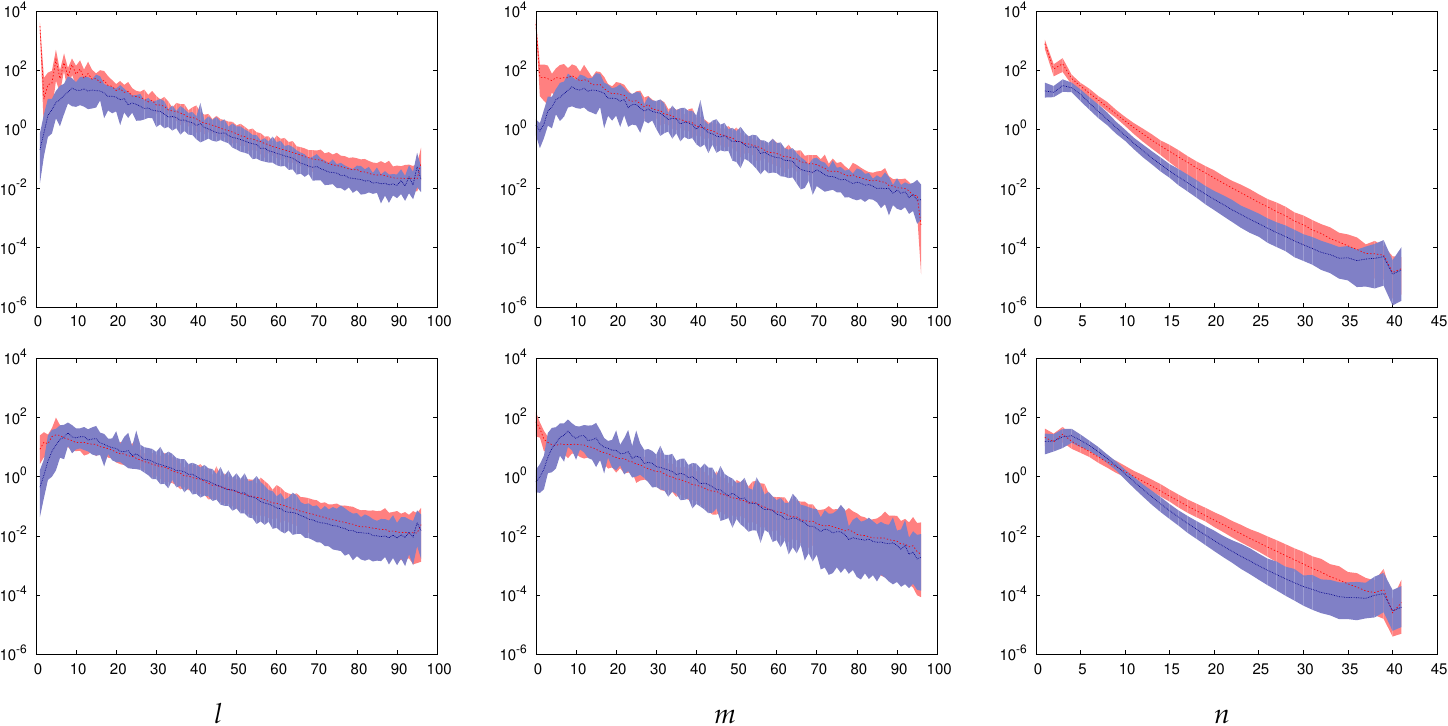}
\end{center}
\caption{Typical power spectra of velocity (blue) and magnetic field  (red).
The top row  shows a \MD{} dynamo solution whereas the bottom row
shows a \FD{} dynamo solution both at $\eta=0.4$, $R=1500000$,
$\tau=2\times10^4$, $P=0.75$ and $P_m=1.5$.
From left to right, power spectra as a function of the spherical
harmonic degree $l$, order $m$, and Chebychev polynomial degree $n$
are shown respectively. Lines represent the average  spectra and
shaded areas go from the minimum to the maximum values for each mode
in the averaging period. A period of one viscous-diffusion time unit
is used for the time-averaging period in both cases.
(color online)}
\label{fig:exampleSpectra}
\end{figure}

\looseness=-1
While typical, the spatial structures described in
relation to figure \ref{fig:3D} are only snapshots of the three dynamo
solutions at fixed  moments in time. An illustration of the
temporal behaviour exhibited in our dynamo simulations is shown in
Figure~\ref{fig:examplePlots}. The main magnetic and kinetic  energy
density components of two distinct dynamo cases are plotted as
functions of time, and the chaotic nature of the solutions is clearly
visible. The time dependence of the time series consist of continual
oscillations around the mean values of the respective densities with
periods much shorter than the viscous diffusion time. 
Kinetic energy densities are displayed in the second row of the figure
and show that the fluctuating components of motion dominate the flow
with the fluctuating toroidal velocity being the strongest. The mean
poloidal component of motion is negligible in both cases in agreement with the
constraint of the Proudman-Taylor theorem on motions parallel to the
axis of rotation. The mean toroidal component, representing
differential rotation, appears to be weak in both cases plotted in
Figure~\ref{fig:examplePlots} more so in the case to the left
marked \MD{} for reasons we will discuss further below. The differential rotation, however is
known to be the component most strongly impaired in the presence of
magnetic field \cite{Simitev2005a}.
This leads us to a discussion of the features of the magnetic energy
densities plotted in the first row of
Figure~\ref{fig:examplePlots}. Here, the differences between the two cases
illustrated are rather more pronounced. The total magnetic energy density of
the case in Figure~\ref{fig:examplePlots}(a) is approximately six
times larger that that in Figure~\ref{fig:examplePlots}(d). More
significant is the essential qualitative difference in the balance of
magnetic energy components. The axisymmetric poloidal component
$\overline{M}_p$\ is dominant in the case shown in  
Figure~\ref{fig:examplePlots}(a) while it has a relatively small 
contribution in the case of Figure~\ref{fig:examplePlots}(d).
The axial dipole coefficient $H_1^0$ and the axial quadrupole
coefficient $H_2^0$ in  Figure~\ref{fig:examplePlots}(c) and (f)
reveal that this difference is due to the fact that the case to the
left is dominated by a strong dipole and the case to the right is 
less strongly dipolar and the time series suggest the presence of
magnetic field oscillations.
\begin{figure}[t]
\begin{center}
\includegraphics[width=0.95\textwidth,clip = true]{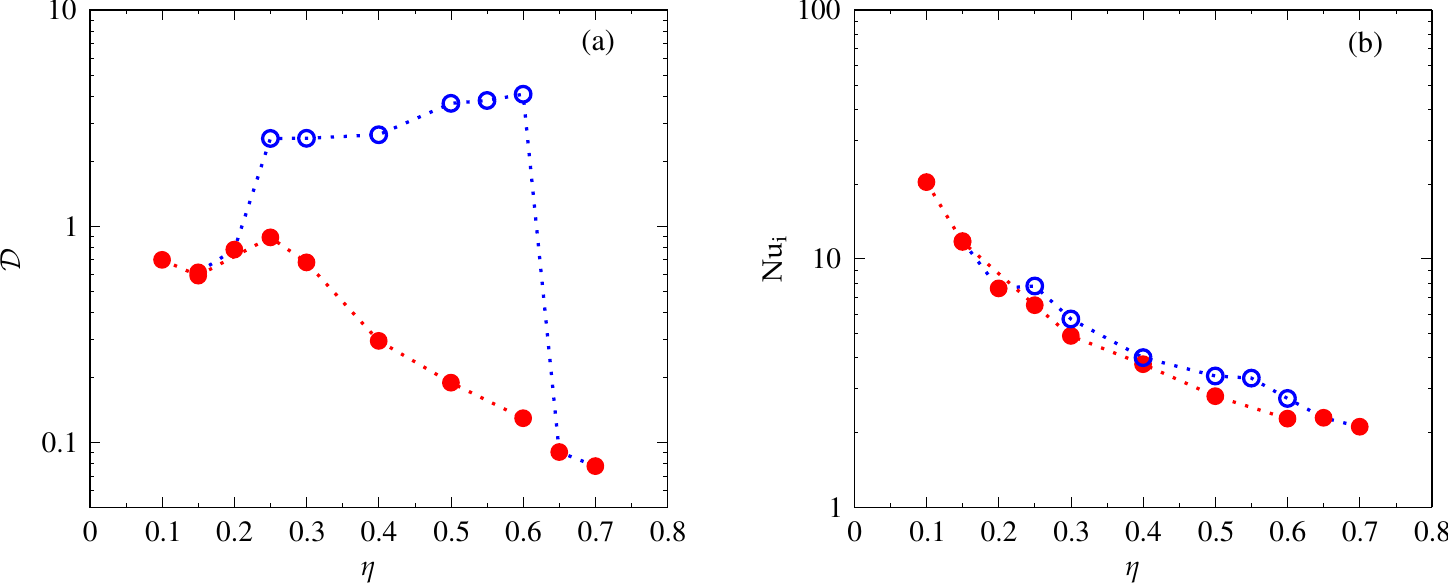}  
\end{center}
\caption{
Bistability as a function of the shell thickness $\eta$. (a) The dipolarity ratio
${\cal D} = \overline{M}_p/\widetilde{M}_p$ and (b) the Nusselt
number at $r=r_i$ in the cases $R=3.8 \times R_c$,  
$\tau=2\times10^4$, $P=0.75$ and $P_m=1.5$. Full red and empty blue circles 
indicate \FD{} and \MD{} dynamos, respectively. Red dotted lines and
blue dotted lines connect dynamos that were started from \FD{} and
\MD{} initial conditions, respectively. 
(color online)}
\label{fig:bistability}
\end{figure}

The solutions plotted in Figure~\ref{fig:examplePlots}(a,b,c) and
\ref{fig:examplePlots}(d,e,f) are  examples of two types  of dipolar
dynamos that have been observed in numerical simulations
\citep{Simitev2005a,Christensen2006,Simitev2009,Olson2011}, namely
those with ${\cal D} > 1$ to which we will refer to as ``Mean
Dipolar'' (\MD{}) and those with ${\cal   D} \leq 1$ that we will call
``Fluctuating Dipolar'' (\FD{}). The typical spatial structures of
the \MD{} and \FD{} dynamos are illustrated in Figure~\ref{fig:exampleMaps}. 
The radial magnetic field plotted in the second column of
Figure~\ref{fig:exampleMaps} shows the predominant dipolar symmetry
of the dynamos, particularly clearly in the \MD{} case where the north and
the south hemispheres have opposite polarities entirely. The \FD{}
case displays a band of reversed polarity in a belt near the equator. In time
this band propagates towards the poles and replaces the initial
polarity leading to a periodically occurring reversals. The
stationary dipole of the \MD{} case is stronger in intensity and
inhibits differential rotation.
This is confirmed by the profiles of the differential rotation plotted
in the left part of the third column of Figure~\ref{fig:exampleMaps}
that are markedly different. The \FD{} case is characterised with a
stronger geostrophic rotation largely aligned with the tangent
cylinder while the mean zonal flow of the \MD{} is weaker and exhibits
a non-geostrophic rotation that is retrograde near the equator. The
columnar convective structure of the solutions remains similar in the
\MD{} and the \FD{} case. Time-averaged kinetic and magnetic energy
power spectra are shown in Figure \ref{fig:exampleSpectra}.

\subsection{Bistability and general effects of shell thickness variation}

One of the most remarkable features of \MD{} and \FD{} dynamos
introduced above is that these two very distinct types can
coexist at identical parameter values. Coexistence was first reported
in \cite{Simitev2009}. Indeed, in each of the Figures
\ref{fig:examplePlots}, \ref{fig:exampleMaps} and
\ref{fig:exampleSpectra} two different cases obtained at the same
parameter values are shown. Within the parameter range of coexistence
it is the initial conditions that determine which of the two chaotic
attractors will be realised.
Figure~\ref{fig:bistability} shows the dipolarity ratio ${\cal D}$  as
a function of the shell thickness aspect ratio $\eta$. Several
observations can be made immediately. First, bistability only seems  
to occur for aspect ratios between $\eta=0.25$ and $\eta=0.6$ and both
to the left and to the right of this interval \FD{} dynamos are
found. In contrast, alternating regimes appeared on each side of the
hysteresis loop in previous studies \cite{Simitev2009,Simitev2012b}
where continuation as function of all remaining parameters $R$, $P$,
$P_m$ and $\tau$ was performed.
A further observation is that the \FD{} dynamos have a decreasing
dipolarity with increasing aspect ratio, that is,  
dipolarity seems to decrease with shell thickness. The \MD{} dynamos, on the other 
hand, show little variation of dipolarity with aspect ratio but can still be 
separated into two groups, one for thin shells and another for thick shells. In 
this respect, it is apparent that thinner shells result dynamos that
are more dipole-dominated.
\begin{figure}[t]
\begin{center}
\includegraphics[width=0.75\textwidth,clip=true]{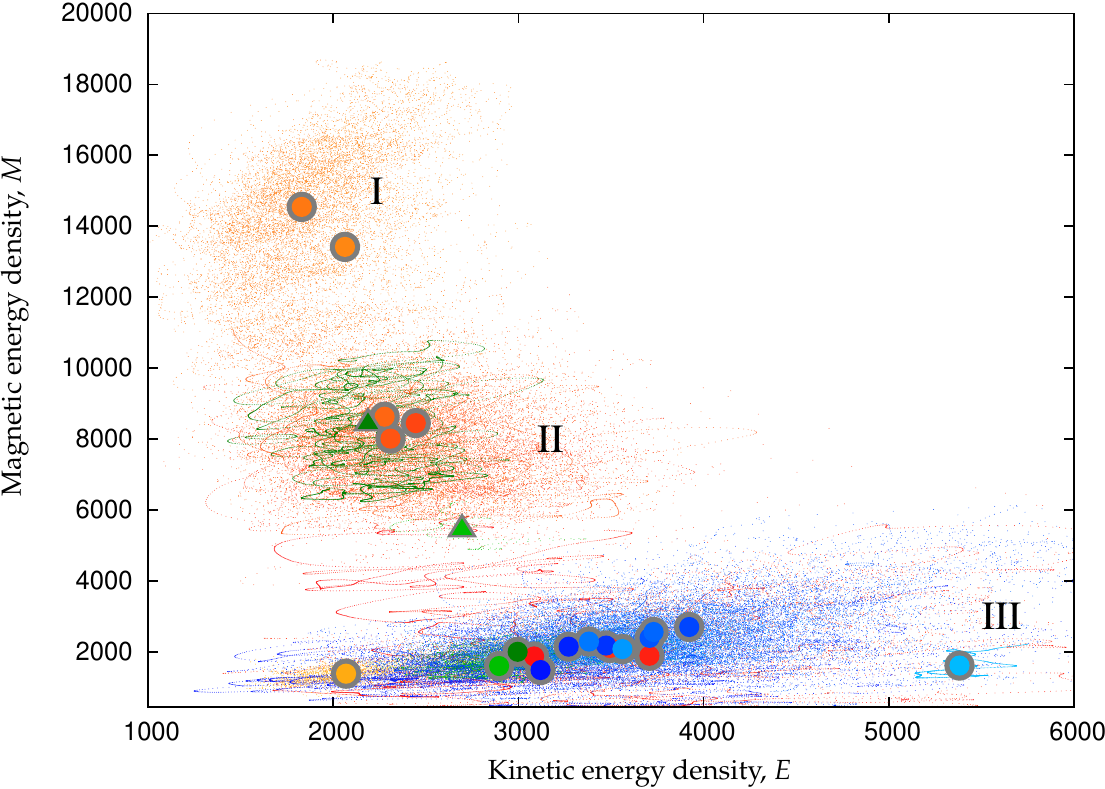}   
\end{center}
\caption{
A ``phase portrait'' of magnetic vs.~kinetic energy density values for
dynamos with $R=3.8 \times R_c$, $\tau=2\times10^4$, $P=0.75$ and
$P_m=1.5$. Dots are instantaneous values; Large markers are time-averaged values. The  
aspect ratio $\eta$ increases from darker to lighter colours (blue to orange). Blue dots and 
points represent dynamos that were started from \FD{} initial
conditions. Warm colours and greens represent simulations that were
started as from \MD{} initial conditions. Green symbols and dots
represent simulations that were
started as from \MD{} initial conditions at $\eta=0.6$\ and
$\eta=0.7$\ and that were repeated starting 
from a higher magnetic energy and lower kinetic energy (triangles) relatively to 
the original simulations (circles).
(color online)}
\label{fig:energyDensityAttractors}
\end{figure}
It is also interesting to note that there is a clear division between \MD{} and \FD{}
dynamos also in the energy density
space. Figure~\ref{fig:energyDensityAttractors} shows a compilation of
plots of  
magnetic energy density as a function of kinetic energy density. Dots represent 
instantaneous values; circles/triangles are mean values over time. The aspect 
ratio, $\eta$, increases from darker to lighter colours. Blue dots and circles 
represent simulations that started off as fluctuating dipolar dynamos whereas warm 
colours and greens represent simulations starting off as mean dipolar
dynamos. Green symbols and dots represent simulations starting off as
mean dipolar dynamos at $\eta=0.6$\ and $\eta=0.7$\ which were
repeated starting  from a higher magnetic energy and lower kinetic
energy (triangles) relatively to  the original simulations (circles).
Three regions can be clearly identified that correspond to simulations that 
finished as high and low dipolarity \MD dynamos (regions I and II in 
Fig.~\ref{fig:energyDensityAttractors}), and to simulations that finished as \FD 
dynamos (region III in Fig.~\ref{fig:energyDensityAttractors}). It is evident 
that dipolarity is preserved  throughout the computations (most warm coloured 
dots and circles end up in region I and II; all blue dots and symbols end up in 
region III). The exception to this rule happens when the magnetic energy density 
of the initial \MD condition is not big enough or its ration to the kinetic 
energy density is small (green circles). In this case the solutions drift to an 
\FD state and remain there. If, on the other hand, the initial \MD condition 
sees its magnetic energy density scaled up sufficiently, the solution will 
remain and \MD dynamo (green dots and triangles).

\begin{figure}[t]
\begin{center}
\includegraphics[width=\textwidth,clip=true]{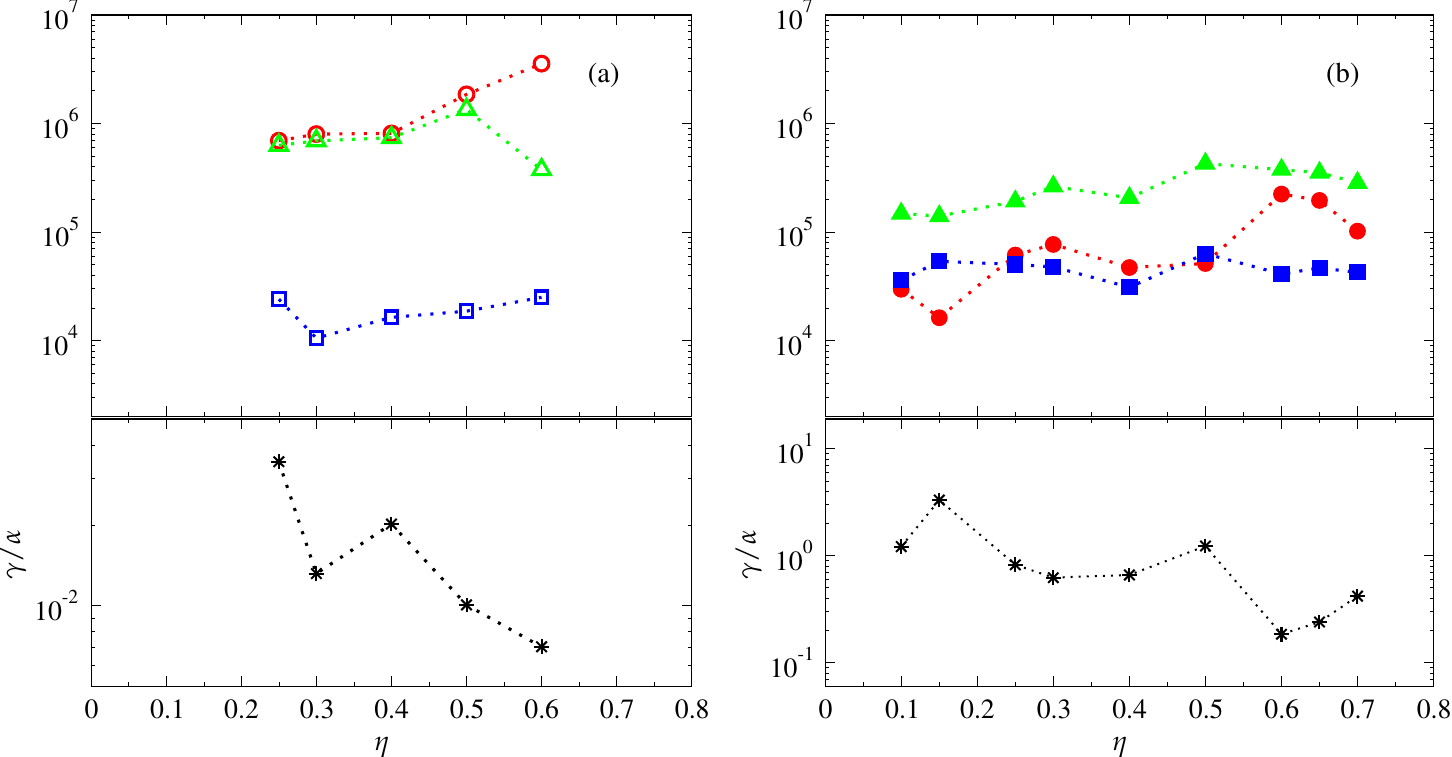}  
\end{center}
\caption{
Magnitude of $\alpha$--, $\beta$--, and $\gamma$--effects with
increasing shell thickness aspect ratio $\eta$ for dynamo solutions with
$R=3.8 \times R_c$, $\tau=2\times10^4$, $P=0.75$ and $P_m=1.5$.
The upper panels show root-mean
squared time-averaged values of the  $\alpha$--effect (red circles), $\beta$--effect
(green triangles up) and $\gamma$--effect (blue squares). The lower
panels  
show the ratio of $\gamma$-- to $\alpha$--effects. Column (a) contains
\MD{} dynamo solutions (empty symbols) while column  (b) contains
\FD{} dynamo solutions (full symbols) as shown in Figure \ref{fig:bistability}. 
(color online)
}
\label{fig:effectsStrength}
\end{figure}

\subsection{The cross-helicity effect}
\label{sec:theCHEffect}
In order to model the effect of turbulence (or, at least, small-scale chaotic motion) on dynamo action, we
consider a separation of scales. This approach is justified as dynamos
tend to exhibit long-lasting large-scale structures (e.g.~the Earth's
dipolar field) together with complex turbulent motions at smaller
scales. We perform an averaging approach where, for the velocity field
$\vec u$ and the magnetic field $\vec b$, we write 
\begin{subequations}
\begin{gather}
 \vec u = \vec U + \vec u', \\
 \vec b = \vec B + \vec b'.
\end{gather}
\end{subequations}
Capital letters represent large-scale components of each field, 
and will be referred to as the ``mean'' components within this and the
following section. As described in the  literature
\cite{KrauseRaedler1980,Yokoi2013,Brandenburg2018}, there are several
ways to perform this scale  separation. Here, we perform the scale
separation by assuming that the steady  large-scale components of the
flow and magnetic field can be identified with  their respective
time-averaged zonal components. The mean flow  is then described as 
\begin{equation}
\vec U = \langle \vec u \rangle = 
     \frac{1}{2\pi\tau}\iint \vec u\ \mbox{d}\varphi\ \mbox{d} t,
     \label{eqn:ind_original}
\end{equation}
for a suitable time scale $\tau$, and a similar expression can be
constructed for the mean magnetic field. In principle, we can apply
this separation of scales to all the main dynamical variables and all
the model equations. Here, however, we only focus on the induction
equation in order to gauge the effect of turbulent transport on the
generation of the magnetic field through dynamo action.   

Applying the above scale separation to the induction equation
\begin{equation}
    \partial_t \vec b =
    \nabla \times \left( \vec u \times \vec b \right)
  + \lambda \nabla^2 \vec b,
\label{eq:mean_mag_ind_eq}
\end{equation}
where $\lambda$ is the magnetic diffusivity (note that equation
\eqref{eq:mean_mag_ind_eq} is an alternative formulation of equation
\eqref{1d}), we find the induction equation for the mean magnetic
field to be
\begin{equation}
\partial_t \vec B =
    \nabla \times \left( \vec U \times \vec B \right)
  + \nabla \times \vec E_\text{M}
  + \lambda \nabla^2 \vec B,
\label{eq:mean_mag_ind_eq}
\end{equation}
where the turbulent electromotive force, $\vec E_\text{M}$, is defined as
\begin{equation}
\vec E_\text{M} = \langle \vec u' \times \vec b' \rangle.
\label{eq:EMF}
\end{equation}
Through an application of the two-scale direct-interaction approximation (TSDIA) of inhomogeneous MHD turbulence (see \cite{Yokoi2020} and references therein), the turbulent electromotive force can be written, in terms of mean variables, as 
\begin{equation}
    \vec E_\text{M} = \alpha\vec B - \beta\vec J  + \gamma \vec \Omega.
\label{eq:EMFEffects}
\end{equation}
Here, $\vec J = \nabla\times\vec B$ and $\vec \Omega =
\nabla\times\vec U$. The coefficients $\alpha$, $\beta$ and $\gamma$
can be expressed in terms of the turbulent residual helicity,
$H=\langle \vec b' \cdot 
\vec j' - \vec u' \cdot \boldsymbol{\omega}'\rangle$,
the turbulent MHD energy,
$K= \langle \vec u'^2 + \vec b'^2\rangle /2$,
and the turbulent cross-helicity
$W=\langle \vec u' \cdot \vec b'\rangle$,
respectively \cite{KrauseRaedler1980,Yoshizawa1990}. Following \cite{Yokoi2013}, they are modelled as
\begin{subequations}
\begin{gather}
\alpha = C_\alpha \tau \langle \vec b' \cdot \vec j' - \vec u' \cdot
\boldsymbol{\omega}'\rangle = C_\alpha \tau H,\\
\beta = C_\beta \tau \langle \vec u'^2 + \vec b'^2 \rangle = C_\beta \tau K,\\
\gamma = C_\gamma \tau \langle \vec u' \cdot \vec b' \rangle = C_\gamma \tau W,
\end{gather}
\end{subequations}
with $C_\alpha$, $C_\beta$ and $C_\gamma$ being model constants. Here, $\tau$ is
the characteristic time of turbulence, which is often expressed as
\begin{equation}
  \tau= K/\epsilon,
\end{equation}
with the dissipation rate of the turbulent MHD energy, $\epsilon$, defined by
\begin{equation}
\epsilon = \nu \left\langle\frac{\partial u'_a}{\partial
  x_b}\frac{\partial u'_a}{\partial
  x_b}\right\rangle +
\lambda \left\langle \frac{\partial b'_a}{\partial
  x_b}\frac{\partial b'_a}{\partial
  x_b}\right\rangle.
\end{equation}
Substituting \eqref{eq:EMFEffects} into the mean induction equation
\eqref{eq:mean_mag_ind_eq}, we have
\begin{equation}
        {\partial_t \vec B}
        = \nabla \times \left( {
                \vec U \times \vec B
        } \right)
        + \nabla \times \left( {
          \alpha \vec B
                + \gamma {\vec{\Omega}}
        } \right)
        - \nabla \times \left[ {
                (\lambda + \beta) \nabla \times \vec B
        } \right].
        \label{eq:mean_mag_ind_eq2}
\end{equation}
Thus, in addition to the transport enhancement or structure
destruction due to turbulence through the enhanced diffusion
$\lambda+\beta$, there is also transport suppression or structure
formation due to turbulence represented by the helicities $\alpha$ and $\gamma$ \cite{Yokoi2020}. 

In the classical mean field theory of dynamos \cite{KrauseRaedler1980,Moffatt1978}, the turbulent
electromotive force is composed of the first two terms on the
right-hand side of equation \eqref{eq:EMFEffects}, namely $\alpha\vec
B-\beta\vec J$. Dynamos resulting from this model are known as
``$\alpha$ dynamos'', where the turbulent diffusion is balanced by an
$\alpha$-effect. The properties of these terms have been discussed
widely in the literature, and so we do not repeat this discussion
here. Instead, let us now consider the final term on the right-hand
side of equation \eqref{eq:EMFEffects}, $\gamma\vec\Omega$. Unlike the other
terms describing the electromotive force, the mean variable in this
term depends on the mean velocity and not the mean magnetic
field. \citet{Yokoi2013} describes how a fluid element subject to a Coriolis-like
force (a mean vorticity field) can contribute to the turbulent
electromotive force through $\gamma$, a measure of the turbulent cross
helicity. Dynamos in which the main balance is between $-\beta\vec J$
and $\gamma\vec\Omega$ are known as ``cross-helicity dynamos'', where
the cross-helicity term replaces the $\alpha$-effect term in balancing
the turbulent diffusion.  
Cross-helicity dynamos have been studied much less than $\alpha$
dynamos, and this study represents an  initial step in addressing this
potentially important imbalance. In particular in Figure
\ref{fig:effectsStrength}, we calculate all three contributions
to the turbulent electromotive force in our dynamo simulations in
order to determine their relative importance. These results are
discussed below.
\begin{figure}[t]
  \begin{center}
\includegraphics[width=\textwidth,clip=true]{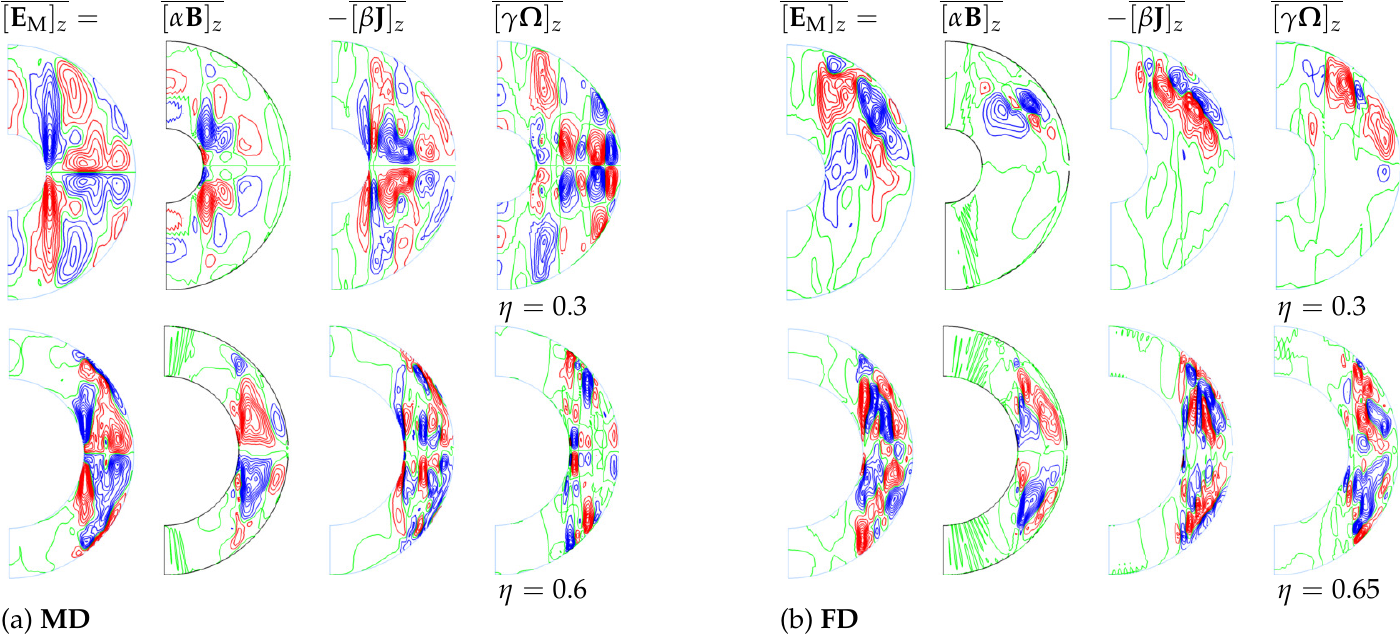}  
\end{center}
\caption{
Spacial structures of the azimuthally-averaged $z$-component
of the electromotive force $\vec E_\text{M}$ and
its $\alpha$--, $\beta$-- and $\gamma$--effect constituents as given
by Equation \eqref{eq:EMFEffects}. Four dipolar dynamo solutions are plotted
as follows.
(a) \MD{} dynamo solutions with
$\eta=0.3$, $P=0.75$, $\tau=2\times10^4$, $R=2500000$, $P_m=1.5$ (top row) and
$\eta=0.6$, $P=0.75$, $\tau=2\times10^4$, $R=410000$, $P_m=1.5$ (bottom row).
(b) \FD{} dynamo solutions with 
$\eta=0.3$, $P=0.75$, $\tau=2\times10^4$, $R=2500000$, $P_m=1.5$ (top row) and
$\eta=0.65$, $P=0.75$, $\tau=2\times10^4$, $R=300000$, $P_m=1.5$ (bottom row).
In each column contour lines of the quantities denoted at the column
heading are plotted with positive contours shown in red, negative
contours shown in blue, and the zeroth contour shown in green.
  (color online) 
}
\label{fig:06md}
\end{figure}

\subsection{Properties and relative importance of cross-helicity}
\label{sec:effectsImportance}

The variation of the turbulent transport coefficients $\alpha$,
$\beta$, and $\gamma$ as a function of shell thickness is displayed in
Figure \ref{fig:effectsStrength}. For simplicity, in this initial
investigation, we take $C_A\tau=1$, where $A=\alpha$, $\beta$, or
$\gamma$. Thus, the three effects are represented
by the turbulent residual helicity $H$, the turbulent MHD energy $K$ and
the turbulent cross-helicity $W$, respectively.
For {\bf MD} dynamo solutions, there is a clear disparity between the
$\alpha$- and $\beta$-effects, and the $\gamma$-effect. The
$\gamma$-effect is, for the range of $\eta$ considered, about two
orders of magnitude smaller than the other effects. Thus, across a wide
range of shell thickness aspect ratios, {\bf MD} dynamos can be considered to be
operating predominantly as $\alpha$ dynamos. 
In contrast, for {\bf FD} dynamo solutions, a different picture
emerges. Across the range of $\eta$ considered, the $\alpha$- and
$\gamma$-effects are of a similar magnitude. Thus, both these effects
are potentially important in balancing the $\beta$-effect. Therefore, {\bf FD}
dynamo solutions represent a ``mixture'' of an $\alpha$ dynamo and a
cross-helicity dynamo.   

Figure \ref{fig:06md} displays $z$-projections of the
azimuthally-averaged components of the electromotive force. For the
\MD{} dynamo solutions, shown in (a), the $\gamma$-effect follows an
antisymmetric pattern about the equator, just like the other
effects. This behaviour is expected from the pseudoscalar nature of
$\gamma$ and the symmetry of magnetic fields in \MD{} dynamos
\cite{Yokoi2013}. For \FD{} dynamo solutions, such as those displayed
in (b), the components of the electromotive force no longer exhibit
antisymmetry about the equator. This behaviour is, in part, due to the
more complex spatial structure of the magnetic fields of \FD{} dynamos
compared to \MD{} dynamos. This feature, combined with generally
weaker magnetic field strengths and different flow profiles (see
Figures \ref{fig:examplePlots} and \ref{fig:exampleMaps}, for
example), results in the $\alpha$-effect being weaker for \FD{}
dynamos. Thus, both the $\alpha$- and $\gamma$-effects become of
comparable importance in sustaining dynamo action.   

\section{{Summary and Discussion}}
\label{sec:disc}

Rotating thermal convection is ubiquitous within the interiors and the atmospheres
of celestial bodies. These fluid regions usually contain plasmas or
metallic components so vigorous convection drives large-scale
electric currents and generate the self-sustained magnetic fields
characteristic of these cosmic objects. In this article the 
relative importance of two  main mechanisms for magnetic field
generation and amplification is assessed, namely the
helicity- and the cross-helicity effects of mean-field dynamo theory.
{The motivation for this study is to test the hypothesis that the
turbulent helicity effect, also known as the $\alpha$--effect, is  more
important in the case of the geodynamo, while the cross-helicity
effect, also known as the $\gamma$-effect, is more significant in the
case of the solar global dynamo, due to differences between the
shell aspect ratio of the solar convection zone and that of Earth's inner
core.}
{The following novel results are reported in the article.
\begin{enumerate}[label=(\alph*),align=left,topsep=0.5ex,parsep=1.5ex]
\item Critical parameter values for onset of convection determined
  numerically as functions of the shell radius ratio, $\eta$. 
\item Bistability and coexistence of two distinct dynamo attractors
  found as a function of the shell radius ratio, $\eta$.   
\item  Spatial distributions and time-averaged values of turbulent
  helicity and cross-helicity EMF effects obtained (1) for both types of
  dynamo  attractors, as well as (2) as functions of the shell radius
  ratio, $\eta$.
\end{enumerate}
Further details and a discussion of these results follows.}

To assess $\alpha$- and $\gamma$- electromotive effects, {we performed,
and report here, an extensive suite of over 40 direct numerical simulations of
self-sustained dynamo action driven by thermal convection in rotating
spherical fluid shells, where the shell thickness aspect ratio $\eta$
is varied at fixed values of the other parameters.}
The simulations are based on the Boussinesq approximation of the
governing non-linear magnetohydrodynamic equations with stress-free
velocity boundary conditions. 
{While the use of fully compressible equations is desirable, it is
not feasible for global dynamo simulations. Indeed, the fully
compressible MHD equations allow sound wave solutions with periods
many orders of magnitude shorter than the convective turnover time and
the magnetic diffusion timescales that are of primary interest.
The Boussinesq approximation is justified and generally used for 
modelling convection in Earth's inner core where density variation
between the inner-outer core boundary and the core mantle boundary is
small \cite{Christensen1999,Roberts2013,Jones2015,Wicht2019}.
The density contrast between the bottom ($\rho_i$) and the top
($\rho_o$) of the Solar convection zone is five orders of magnitude
giving a density scale number of $\log(\rho_i/\rho_o)\approx12$
\cite{ChristensenDalsgaard1996}, and the anelastic approximation is   
more appropriate and commonly used in global solar convection models,
e.g.~\cite{Charbonneau2014,Simitev2015,Matilsky2020}. However,   
anelastic and Boussinesq simulations show many similarities
\cite{Simitev2015}, with Boussinesq models able to mimic solar periodicity
and active longitude phenomena \cite{Simitev2012a,Simitev2012b}. Thus,
in this work the Boussinesq approximation is used for uniformity
across various shell radius ratios and to focus on the effects of shell
thickness in isolation from effects of density stratification.}

Coexistence of distinct chaotic dynamo states has been
reported to occur in this problem in terms of certain
governing parameters in \cite{Simitev2009,Simitev2012c}. {In this study, we
establish that two essentially different nonlinear dynamo attractors
coexist also for an extensive range of shell thickness aspect ratios
$\eta\in[0.25,0.6]$.
Since this is precisely the range of values where most celestial
dynamos operate this result is significant as it demonstrates that
field morphologies may be dependent on the initial state of a dynamo. 
We proceed to discuss in detail the
contrasting properties characterizing the coexisting dynamo regimes
(mean-field dipolar (\MD{}) dynamos and fluctuating dipolar (\FD{}) dynamos) including
differences in temporal behavior and spatial structures of both the
magnetic field and rotating thermal convection.
We find that the relative importance of the electromotive dynamo
effects is different in the cases of mean-field dipolar dynamos and
fluctuating dipolar dynamos.
The helicity $\alpha$-effect and the cross-helicity $\gamma$-effect
are comparable in intensity in the case of fluctuating dipolar
dynamos and their ratio does not vary significantly with shell thickness.
In contrast, in the case of mean-field dipolar dynamos the helicity
$\alpha$-effect dominates by approximately two orders of magnitude and
becomes even stronger with decreasing shell thickness.}
Our results, therefore, indicate that both dynamo mechanisms are
important for solar global magnetic field generation as the solar
dynamo is of a fluctuating dipolar type. Our results also indicate
that the cross-helicity effect may be important in understanding
dynamo mechanisms in stellar dynamos. The latter may also be of fluctuating
dipolar type and markedly different from the solar dynamo, e.g. having 
large-scale magnetic structures being dominant in only one hemisphere
\cite{MacTaggart2016}.  Since the geodynamo is of a 
mean-field dipolar type, the helicity effect appears, indeed, to be
more significant in this case and our results show this effect will
become even stronger as the inner solid core grows in size by iron
freezing.
{Simulations of the geodynamo with nucleation and growth
of the inner core have been recently reported by \citet{Driscoll2016}
and \citet{Landeau2017}. These authors find that 
pre-inner core nucleation dynamos exhibit weak thermal convection, low
magnetic intensity and non-dipolar field morphology, while post-inner
core nucleation and with increasing inner core size their solutions 
have stronger axial dipole morphology. Our results similarly
demonstrate that \FD{} and multipolar dynamos  
occur when the value of the shell radius ratio $\eta$ is smaller than
0.25. However, our \FD solutions exhibit 
vigorous convection and can be described as strong-field dynamos even
though of lower magnetic field intensity than corresponding \MD{}
dynamos. A further discrepancy is that for $\eta>0.25$ we find that
\MD{} and \FD{} dynamos coexist. These discrepancies can be attributed
to significant differences in thermal and velocity boundary conditions
between our model and the models of \cite{Driscoll2016,Landeau2017}.
Most importantly, the governing parameters values in
\cite{Driscoll2016,Landeau2017} are controlled by thermochemical evolution
models and vary with inner core size (age), while in our study all
parameter values apart from $\eta$ are kept fixed.}  

It will be of interest to revisit {the analysis of helicity and
cross-helicity effects using the more general anelastic approximation
of the governing equations}. Further, there are
many questions that remain to be answered on how the dynamic balance
between the components of the electromotive force affects different
aspects of dynamo action, including how to switch between \MD{} and
\FD{} dynamos.

\authorcontributions{
Conceptualization, R.S; methodology, R.S.; software, R.S.; validation,
L.S.; formal analysis, R.S. and D.MT.; investigation, L.S. and P.G.;
resources, R.S.; data curation, L.S. and P.G.; writing--original draft
preparation, R.S.; writing--review and editing, R.S. and D.MT.;
visualization, L.S, R.S. and P.G.; supervision, R.S.; funding acquisition, R.S.}

\funding{This research was funded by the Leverhulme Trust grant number RPG-2012-600.}

\acknowledgments{Numerical simulations were carried out in part at the
DiRAC Data Centric system at Durham University,  
operated by the Institute for Computational Cosmology on behalf of the
STFC DiRAC HPC Facility (www.dirac.ac.uk). This equipment was funded
by BIS National E-infrastructure capital grant ST/K00042X/1, STFC
capital grants ST/H008519/1 and ST/K00087X/1, STFC DiRAC Operations
grant ST/K003267/1 and Durham University. DiRAC is part of the
National E-Infrastructure.}

\conflictsofinterest{The authors declare no conflict of interest.}

\abbreviations{The following abbreviations are used in this manuscript:\\

\noindent 
\begin{tabular}{@{}ll}
\MD{} & Mean Dipolar Dynamo\\
\FD{} & Fluctuating Dipolar Dynamo\\
\end{tabular}}


\reftitle{References}





\end{document}